\documentclass[sigconf]{acmart}

\AtBeginDocument{%
  \providecommand\BibTeX{{%
    \normalfont B\kern-0.5em{\scshape i\kern-0.25em b}\kern-0.8em\TeX}}}

\copyrightyear{2021}
\acmYear{2021} 
\setcopyright{iw3c2w3}
\acmConference[WWW '21]{Proceedings of the Web Conference 2021}{April 19--23, 2021}{Ljubljana, Slovenia} 
\acmBooktitle{Proceedings of the Web Conference 2021 (WWW '21), April 19--23, 2021, Ljubljana, Slovenia}
\acmPrice{}
\acmDOI{10.1145/3442381.3449866}
\acmISBN{978-1-4503-8312-7/21/04}

\usepackage{amsthm}
\newtheorem{definition}{Definition}
\usepackage{color}
\usepackage{multirow}
\usepackage{multicol}
\usepackage{graphicx}
\usepackage{subcaption}
\usepackage{pgfplots}\pgfplotsset{compat=1.9}
\usepackage[skip=5pt]{caption}

\definecolor{celadon}{rgb}{0.67, 0.88, 0.69}
\definecolor{darkseagreen}{rgb}{0.56, 0.74, 0.56}
\begin{document}

%%
%% The "title" command has an optional parameter,
%% allowing the author to define a "short title" to be used in page headers.
% \title{User-oriented Group Fairness in Recommender Systems}
\title{User-oriented Fairness in Recommendation}

%%
%% The "author" command and its associated commands are used to define
%% the authors and their affiliations.
%% Of note is the shared affiliation of the first two authors, and the
%% "authornote" and "authornotemark" commands
%% used to denote shared contribution to the research.

% \settopmatter{authorsperrow=3}

% \author{Yunqi Li, Hanxiong Chen, Zuohui Fu, Yingqiang Ge and Yongfeng Zhang}
% \affiliation{
%   \institution{Department of Computer Science, Rutgers University, New Brunswick, NJ 08901, US}
% }
% \email{{yunqi.li,hanxiong.chen,zuohui.fu,yingqiang.ge,yongfeng.zhang}@rutgers.edu}

\author{Yunqi Li}
\affiliation{
  \institution{Rutgers University}
  \city{New Brunswick, NJ, US}
}
\email{yunqi.li@rutgers.edu}

\author{Hanxiong Chen}
\affiliation{
  \institution{Rutgers University}
  \city{New Brunswick, NJ, US}
}
\email{hanxiong.chen@rutgers.edu}

\author{Zuohui Fu}
\affiliation{
  \institution{Rutgers University}
  \city{New Brunswick, NJ, US}
}
\email{zuohui.fu@rutgers.edu}

\author{Yingqiang Ge}
\affiliation{
  \institution{Rutgers University}
  \city{New Brunswick, NJ, US}
}
\email{yingqiang.ge@rutgers.edu}

\author{Yongfeng Zhang}
\affiliation{
  \institution{Rutgers University}
  \city{New Brunswick, NJ, US}
}
\email{yongfeng.zhang@rutgers.edu}

%%
%% By default, the full list of authors will be used in the page
%% headers. Often, this list is too long, and will overlap
%% other information printed in the page headers. This command allows
%% the author to define a more concise list
%% of authors' names for this purpose.
% \renewcommand{\shortauthors}{Trovato and Tobin, et al.}

%%
%% The abstract is a short summary of the work to be presented in the
%% article.
\begin{abstract}
As a highly data-driven application, recommender systems could be affected by data bias, resulting in unfair results for different data groups, which could be a reason that affects the system performance. Therefore, it is important to identify and solve the unfairness issues in recommendation scenarios.

In this paper, we address the unfairness problem in recommender systems from the user perspective. We group users into advantaged and disadvantaged groups according to their level of activity, and conduct experiments to show that current recommender systems will behave unfairly between two groups of users. Specifically, the advantaged users (active) who only account for a small proportion in data enjoy much higher recommendation quality than those disadvantaged users (inactive). Such bias can also affect the overall performance since the disadvantaged users are the majority.  To solve this problem, we provide a re-ranking approach to mitigate this unfairness problem by adding constraints over evaluation metrics. The experiments we conducted on several real-world datasets with various recommendation algorithms show that our approach can not only improve group fairness of users in recommender systems, but also achieve better overall recommendation performance. 
\end{abstract}

%%
%% The code below is generated by the tool at http://dl.acm.org/ccs.cfm.
%% Please copy and paste the code instead of the example below.
%%
\begin{CCSXML}
<ccs2012>
   <concept>
       <concept_id>10002951.10003317.10003347.10003350</concept_id>
       <concept_desc>Information systems~Recommender systems</concept_desc>
       <concept_significance>500</concept_significance>
       </concept>
   <concept>
       <concept_id>10010147.10010178</concept_id>
       <concept_desc>Computing methodologies~Artificial intelligence</concept_desc>
       <concept_significance>500</concept_significance>
       </concept>
 </ccs2012>
\end{CCSXML}

\ccsdesc[500]{Information systems~Recommender systems}
\ccsdesc[500]{Computing methodologies~Artificial intelligence}

%%
%% Keywords. The author(s) should pick words that accurately describe
%% the work being presented. Separate the keywords with commas.
\keywords{Recommendation System; Fairness; Re-ranking; AI Ethics}

%% A "teaser" image appears between the author and affiliation
%% information and the body of the document, and typically spans the
%% page.
% \begin{teaserfigure}
%   \includegraphics[width=\textwidth]{sampleteaser}
%   \caption{Seattle Mariners at Spring Training, 2010.}
%   \Description{Enjoying the baseball game from the third-base
%   seats. Ichiro Suzuki preparing to bat.}
%   \label{fig:teaser}
% \end{teaserfigure}

%%
%% This command processes the author and affiliation and title
%% information and builds the first part of the formatted document.
\maketitle

\section{Introduction}
% Recommender systems are highly data-driven and could be negatively influenced by the biases embedded in the data, which may unfairly discriminate among different data groups in terms of recommendation performance \cite{abdollahpouri2017controlling,fu2020fairness}. Therefore, it is important to analyze the potential fairness risks under recommendation scenario, and figure out how to quantify and solve the unfair issue.

% 

% Machine learning, especially deep learning based models are highly data-driven methods which could be affected by the biases embeded in the data. A biased model would generate unfair results to different data groups, which would also be a possible reason for performance degradation \zfcomment{add citation}. As one of the most popular applications of machine learning, recommender systems are potentially affected by the unbalanced data, and thus discriminate unfairly among different data groups in terms of recommendation performance \cite{fu2020fairness}. Therefore, it is important to analyze the potential fairness risks under recommendation scenario, and figure out how to quantify and solve the unfair issue.

\begin{figure}[t]
        \centering
        \pgfplotsset{label style={font=\Large},
                    tick label style={font=\Large}}

        \begin{subfigure}[b]{0.235\textwidth}
            \centering
            \resizebox{1\textwidth}{!}{
                \begin{tikzpicture}
                    \begin{axis}[
                        width  = 2*\textwidth,
                        major x tick style = transparent,
                        ybar=2*\pgflinewidth,
                        bar width=10pt,
                        ymajorgrids = true,
                        ylabel = {F1@10},
                        symbolic x coords={BiasedMF,NeuMF,STAMP},
                        xtick = data,
                        scaled y ticks = false,
                        enlarge x limits=0.25,
                        ymin=0,
                        ymax=40,
                        legend image code/.code={%
                        \draw[#1, draw=none] (0cm,-0.1cm) rectangle (0.6cm,0.1cm);
                        },  
                        legend style={
                            draw=none, % ?
                            text depth=0pt,
                            at={(-0.08,-0.15)},
                            anchor=north west,
                            legend columns=-1,
                            font=\fontsize{11}{5}\selectfont,
                            % default spacing:
                            column sep=0.1cm,
                            % The text "Legend:"
                            % /tikz/column 2/.style={column sep=0pt,font=\bfseries},
                            %
                            % the space between legend image and text:
                            /tikz/every odd column/.append style={column sep=0cm},
                        }
                    ]
                        \addplot[style={celadon,fill=celadon,mark=none}]
                             coordinates {(BiasedMF,36.48) (NeuMF,36.02) (STAMP,33.52)};
                        \addplot[style={red,fill=red!30,mark=none}]
                             coordinates {(BiasedMF,13.86) (NeuMF,12.51) (STAMP,12.57)};
                        \addplot[style={blue,fill=blue!30,mark=none}]
                            coordinates {(BiasedMF, 15.78) (NeuMF,14.51) (STAMP,14.35)};
                        \addlegendentry{Advantaged}
                        \addlegendentry{Disadvantaged}
                        \addlegendentry{Overall}
                    \end{axis}
                \end{tikzpicture}
            }
            \caption[]%
            {{\small Original}}    
            \label{fig:diff_overview_a}
        \end{subfigure}
        \hfill
        \begin{subfigure}[b]{0.235\textwidth}  
            \centering 
            \resizebox{1\textwidth}{!}{
                \begin{tikzpicture}
                    \begin{axis}[
                        width  = 2*\textwidth,
                        major x tick style = transparent,
                        ybar=2*\pgflinewidth,
                        bar width=10pt,
                        ymajorgrids = true,
                        ylabel = {F1@10},
                        symbolic x coords={BiasedMF,NeuMF,STAMP},
                        xtick = data,
                        scaled y ticks = false,
                        enlarge x limits=0.25,
                        ymin=0,
                        ymax=30,
                       legend image code/.code={%
                        \draw[#1, draw=none] (0cm,-0.1cm) rectangle (0.6cm,0.1cm);
                        },  
                        legend style={
                            draw=none, % ?
                            text depth=0pt,
                            at={(-0.08,-0.15)},
                            anchor=north west,
                            legend columns=-1,
                            font=\fontsize{11}{5}\selectfont,
                            % default spacing:
                            column sep=0.1cm,
                            % The text "Legend:"
                            % /tikz/column 2/.style={column sep=0pt,font=\bfseries},
                            %
                            % the space between legend image and text:
                            /tikz/every odd column/.append style={column sep=0cm},
                        }
                    ]
                        % \addlegendimage{empty legend}
                        % \addlegendentry{\textbf{Legend:}}  
                        
                        \addplot[style={celadon,fill=celadon,mark=none}]
                             coordinates {(BiasedMF,21.37) (NeuMF,20.81) (STAMP,19.23)};
                        \addlegendentry{Advantaged}
                        \addplot[style={red,fill=red!30,mark=none}]
                             coordinates {(BiasedMF,15.65) (NeuMF,14.99) (STAMP,14.03)};
                        \addlegendentry{Disadvantaged}
                        \addplot[style={blue,fill=blue!30,mark=none}]
                            coordinates {(BiasedMF, 16.13) (NeuMF,15.48) (STAMP,14.47)};
                        \addlegendentry{Overall}
                        % \legend{No vectorization,TreeScore $>2$,TreeScore $>3$,TreeScore $>4$}
                        % \addlegendimage{my legend}
                        % \addlegendentry{My line}
                        %\draw (rel axis cs: 0,0.3) -- (rel axis cs: 1, 0.3);
                    \end{axis}
                \end{tikzpicture}
            }
            \caption[]%
            {{\small Fair Method}}    
            \label{fig:diff_overview_b}
        \end{subfigure}
        \vspace{-15pt}
        \caption[]{\small (a) Original results show the significant performance difference between advantaged and disadvantaged groups in F1@10; (b) Fairer performance between two groups and better overall performance achieved from our fair re-ranking method. The results here were obtained by controlling the difference of F1@10 between the two groups by less than a quarter of the original.} 
        \label{compare}
\vspace{-15pt}
\end{figure}
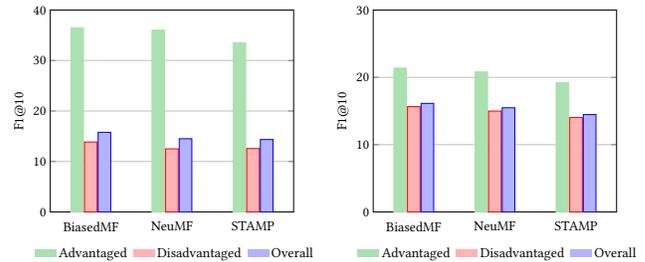

Recently, there has been growing attention on fairness considerations in machine learning community, including classification tasks \cite{zemel2013learning,woodworth2017learning, narasimhan2018learning,cotter2019two} and ranking tasks \cite{ singh2018fairness,zehlike2020reducing,biega2018equity,singh2019policy}, etc. Recommendation algorithms can usually be considered as a type of ranking algorithm. However, the ranking problem usually only considers fairness issue from the perspective of items, while the concept of fairness in recommender systems has been extended to multiple stakeholders \cite{burke2017multisided}, i.e., the unfairness issue should be considered not only from items or providers side, but also ought to be taken care of from the user side. Comparing with the sufficient work about solving the discrimination from items side in recommendations \cite{adomavicius2011improving,kamishima2014correcting,abdollahpouri2017controlling,abdollahpouri2019managing}, algorithmic bias existing on the user side has been rarely studied. 

% In this paper, we aim at mitigating the recommendation performance bias from the user perspective by re-ranking the recommendation lists generated by a base recommendation algorithm, to take the advantage of making no assumption of the underlying recommendation model, and offering the flexibility to be applied to many different models.

In this paper, we consider unfairness issues between different group of users regarding recommendation performance in commercial recommendation scenarios. The unfairness could result from the data imbalance. 
Researchers have shown that recommender systems may suffer from the item popularity bias, i.e., the popular items can get more exposure than those unpopular ones in recommendations \cite{park2008long}. The underlying mechanism is that popular items will gain more visibility when training the recommendation model due to their sufficient data, and thus the model may be biased to or even dominated by these items. 

The data imbalance and algorithmic bias problem that exists on the item side also exists on the user side. Specifically, users who interact with the platform more actively will contribute more sufficient data than those less active users when training the model. Due to the fundamental idea of collaborative filtering in most recommendation algorithms, this would lead to the problem that the trained recommender systems would be biased towards or even dominated by those more active users.
% , since it is easier to capture their preference in model training due to their sufficient data. 
As a result, the users with less activity are more likely to receive unsatisfied recommendation results. This will give rise to the unfair treatment between user groups with different activity levels. Here we call such recommendation performance disparity between more and less active user groups as unfairness, since it is caused by the bias that exists in data and the algorithmic bias in some recommendation algorithms. What's more, such unfairness issue can also be a reason of the system's overall performance degradation since those less active users play the majority portion in most cases. Therefore, it is essential to pay attention to the majority of the less active users so that they can enjoy more satisfactory recommendation experience, and thus improve the overall recommendation quality.
% We call such recommendation performance disparity between more and less visible user groups unfairness since it is caused by the bias which exists in data, and even the algorithmic bias in some recommendation algorithms. 
% This can also be a reason of systems overall performance degradation since those less visible users are in the majority in most cases. 
% % With these concerns, we believe that user-level fairness research is worth working on. However, such unfairness of recommendations from a user's visibility perspective is rarely studied.
% With this concerns, we divide users by their interactions into different groups, and we call the group of users with higher interactions visibility advantaged group, while the remaining users visibility disadvantaged groups. We also provide a framework to mitigate such algorithmic unfairness issue in this paper.

% Affected by the fundamental collaborative filtering idea, current recommender systems are easy to be dominated by the data of users who actively interact with platforms. However, such visible users may only account for a very small proportion. Therefore, it is important to also pay attention to the majority less visible users so that they can enjoy the same recommendation experience, and thus improve the overall recommendation quality. 

To capture users' activity level, we explore three grouping methods through observable information to distinguish users into different activity level, including their number of interactions; total consumption capacity (i.e., the accumulative price the user consumed); and maximum consumption level (i.e., the maximum price of items that the user bought). We label those more active users as advantaged group, while the remaining users as disadvantaged group. The reason for exploring these three methods is that we believe the difference of user interactions and consumption power will reflect their different activity level in a reasonable manner, since usually users who interact with the e-commerce platform more actively will tend to make more purchases, show
higher consumption capacity, and have greater consumption budget. We conduct data-driven analysis to explore the performance of some shallow or deep recommender systems on several Amazon datasets under the three grouping methods. Specifically, we discover that the user distribution is more concentrated in the area with fewer interactions or lower consumption, while the advantaged users---who only account for a very small proportion of users in data---enjoy significantly higher recommendation quality than those disadvantaged users, as well as the overall recommendations performance. To solve such unfairness issue, we provide a framework based on re-ranking with fairness constraints to mitigate the performance disparity.

In summary, we aim at mitigating the unfairness of recommendations from the user perspective in this paper. We differentiate users into advantaged and disadvantaged groups in commercial recommendation systems according to their level of activity, and find that disadvantaged users are more likely to receive unsatisfied recommendations because of their insufficient training data compared with advantaged users. To address the unfairness problem above, we provide a re-ranking method with user-oriented group fairness constrained on the recommendation lists generated from any base recommender algorithm. The re-ranking strategy helps to mitigate the recommendation performance bias by taking advantage of making no assumptions of the underlying recommendation model, and offering the model-agnostic
flexibility. 
% to be applied to many different models.
Our experiments on three Amazon datasets with different types of shallow or deep recommendation algorithms show that our method is not only able to reduce unfairness between two user groups, but also improve the overall recommendation performance. Figure \ref{compare}(a) shows the significant algorithmic unfairness of the recommendations between advantaged and disadvantaged groups, and Figure \ref{compare}(b) shows the results generated by our re-ranking method, which provides fairer and best overall recommendation performance.

The key contributions of this paper are as follows:
\begin{itemize}
    \item We state the importance of concerning the unfairness issue caused by data imbalance between user groups with different activity level in commercial recommendation systems. We explore three methods to capture user activity levels using observable information.
    
    % in recommender systems, and propose three methods to distinguish users into advantaged and disadvantaged groups with observable information.
    % % % \item Different from most of the previous researches which solve the item popularity bias in recommendations, we consider unfairness recommendation from the perspective of user visibility. We conduct a data-driven observational analysis to show the performance of four traditional recommendation algorithms on three Amazon datasets to argue that such algorithms produce unfair recommendations between different group of users, and thus reduces the overall recommendation performance.
    % \item We show that recommender systems could suffer from generating biased results based on our grouping strategy. 
    \item We provide a fairness constrained re-ranking method and formalize it as a 0-1 integer programming problem to reduce the bias.
    \item We conduct extensive experiments on three Amazon datasets with four shallow or deep recommendation algorithms to show that our method can shrink not only the fairness disparity between different groups of users, but also improve the overall recommendation quality.
    
\vspace{-1.2pt}   
\end{itemize}

In the following, we review related work in Section \ref{sec:related} and motivate the fairness concerns in Section \ref{sec:motivation}. In Section \ref{sec:framework}, details of our framework are introduced. Experimental settings and results are provided in Section \ref{sec:experiments}. Finally, we conclude this work in Section \ref{sec:conclusions}.

% Addressing unfairness in intelligent decision making systems has become an increasingly important problem due to their wider application. 

%  Recommender systems, as one of the most pervasive applications of decision making systems in industry, may suffer from performance disparity

% It is important to identify the potential fairness risks in recommender systems and address the unfairness. 

% in recommender systems has become an increasingly important problem due to the growing influence of rankings in critical decision making, yet existing recommendation algorithms suffer from unfair

% Recommender systems are one of the most pervasive applications of machine learning in industry,
% with many services using them to match users to products or information. 

% There has been growing attention on fairness considerations recently, especially in the context of intelligent decision making systems.

\section{Related Work}
\label{sec:related}
\subsection{Algorithmic Fairness}
Fairness is becoming one of the most important topics in machine learning in recent years \cite{kamishima2012fairness,pedreshi2008discrimination, singh2018fairness}. There are two basic frameworks adopted in recent studies on algorithmic discrimination: individual fairness and group fairness. Individual fairness requires that each similar individual should be treated similarly, which is hard to define precisely due to the lack of agreement on task-specific similarity metrics for individuals \cite{dwork2012fairness}. Group fairness requires that the protected groups should be treated similarly to the advantaged group or the populations as a whole \cite{pedreschi2009measuring}. The group fairness perspective for supervised learning usually implies constraints such as equalized odds and demographic parity. Equalized odds defines the constraint that the false positive rate and true positive rate should be equal for the protected group and advantaged group, which represents the equal opportunity principle \cite{hardt2016equality,zafar2017fairness}. Demographic parity, also called independence or statistical parity, is one of the most well-known criteria for fairness \cite{calders2009building}. It requires that decisions should be similar around a sensitive attribute such as gender or nationality. The flaw is that demographic parity will cause a loss in the utility and also infringes individual fairness \cite{dwork2012fairness}. Most recent works about fairness concerns have focused on designing algorithms compatible with such fairness constraints on fair classification \cite{zemel2013learning,woodworth2017learning}. The fairness metrics for binary classification problems can be written in terms of rate constraints, which are on the classiﬁer's positive or negative prediction rate for different protected groups \cite{narasimhan2018learning,cotter2019two}. For example, demographic parity posits that the classiﬁer's positive prediction rate is the same across all groups. Such constraints for fairness metrics can be added to the training objective for a binary classiﬁer, and be solved using constrained optimization algorithms or relaxation methods \cite{goh2016satisfying,agarwal2018reductions}. Here, we address group unfairness in recommender systems from user perspective.

\subsection{Fair Ranking}
Besides fairness concerns in classification, some recent works have raised the question of fairness in rankings. Recommendation algorithms can usually be considered as a type of ranking algorithm. However, existing work usually only consider fairness issue from the item perspective in ranking problem, while the concept of fairness in recommender systems is more complicated as the unfair issue will also lie on the user side.

\citeauthor{biega2018equity} \cite{biega2018equity} capture unfairness at the level of individual subjects, as such subsume group unfairness. They claim that no single ranking can achieve individual attention fairness, so they propose a new mechanism to quantify and mitigate the position bias,  which leads to disproportionately less attention being paid to low-ranked subjects. The results achieve amortized fairness by making the attention accumulated across a series of rankings to be proportional to accumulated relevance. Nowadays, most existing works measure unfairness in ranking at the level of subject groups. The fairness metrics are usually relevant to the exposure of the items belonging to a different protected group. As concluded in \cite{narasimhan2020pairwise}, the metrics include the supervised criteria which control the average exposure of groups to be proportional to the average relevance of the results of groups to a query \cite{biega2018equity,singh2019policy}, and the unsupervised criteria which require that the average exposure at the top of the ranking list is equal for different groups \cite{celis2017ranking, singh2018fairness,zehlike2020reducing}. In a fair ranking problem, some research works directly learn a ranking model from scratch \cite{singh2019policy,zehlike2020reducing,narasimhan2020pairwise}, while others consider re-ranking or post-processing algorithms after a ranking has been given \cite{celis2017ranking, biega2018equity}. In this paper, we use re-ranking method to reduce discrimination to take advantage of making no assumptions to the underlying recommendation models and offering the flexibility to be applied to different models.
\vspace{-20pt}
\subsection{Fair Recommendation}
There has been a small amount of work on fairness in recommendation task, and each work takes very different perspectives. Different from fair ranking and classification, in the field of recommendation systems, the concept of fairness has been extended to multiple stakeholders \cite{burke2017multisided}. The unfair issue can be considered not only from the item or the provider side, but also can be considered from the user side, which makes the problem to be more complex. Both \cite{burke2017multisided} and \cite{abdollahpouri2019multi} categorize different types of multi-stakeholder platforms and the different group fairness properties they desired. \citeauthor{mehrotra2018towards} \cite{mehrotra2018towards} address the supplier fairness in two-sided marketplace platforms and propose a heuristic strategy to jointly optimize fairness and performance. \citeauthor{patro2020fairrec} \cite{patro2020fairrec} address individual fairness for both producers and customers, and answer the question of the long-term sustainability of two-sided platforms. \citeauthor{yao2017beyond} \cite{yao2017beyond} study fairness in collaborative filtering recommender systems,  and propose four new metrics that address different forms of unfairness. These fairness metrics can be optimized by adding fairness terms to the learning objective. \citeauthor{xiao2017fairness} \cite{xiao2017fairness} provide an optimization framework for fairness-aware group recommendation from the perspective of Pareto Efficiency, and further explore the fairness of measure trade-off in recommender systems under a Pareto optimization framework \cite{lin2019pareto}. \citeauthor{beutel2019fairness} \cite{beutel2019fairness} show how to measure fairness based on pairwise comparisons from randomized experiments, and offer a regularizer to improve fairness when training recommendation models. \citeauthor{leonhardt2018user} \cite{leonhardt2018user} quantify the user unfairness caused by the post-processing algorithms which have the original goal of improving diversity in recommendations. \citeauthor{ge2021towards}\cite{ge2021towards} explore long-term fairness in recommendation and accomplish the problem through dynamic fairness learning. \citeauthor{fu2020fairness} \cite{fu2020fairness} propose a fairness constrained approach to mitigate the unfairness problem in the context of explainable recommendation over knowledge graphs. They find that performance bias exists  between different user groups, and claim that such bias comes from the different distribution of path diversity. Here, we show that such recommendation performance bias also exists in general recommender systems. There are more researches concerning the popularity bias problem in recommendations, i.e., the frequently rated items will get more exposure than those less popular ones. Such researches mainly solve this problem by increasing the number of recommended unpopular items (long-tail items) or otherwise the overall catalog coverage \cite{adomavicius2011improving,kamishima2014correcting,abdollahpouri2017controlling,abdollahpouri2019managing}.
\citeauthor{abdollahpouri2019unfairness} \cite{abdollahpouri2019unfairness} see the problem from the users’ perspective with finding how popularity bias causes the recommendations to deviate from what the user expects to get from the recommender system. In this paper,  we concern about the unfair issue caused by the bias on the user side, and  reasonably divide users into different groups according to their behaviour \cite{li2016graph,li2016art}.

\section{MOTIVATING FAIRNESS CONCERNS}
\label{sec:motivation}
\label{MFC}
\begin{table}[t]
  \setlength{\tabcolsep}{4pt}
  \caption{Percentage of users located at different number of interactions thresholds (as $n$ represents) in the training set of the datasets.}
  \label{tab:freq}
  \begin{tabular}{lllll}
    \toprule
    Dataset& $n\geq$5 &$n\geq$10  & $n\geq$20& $n\geq$30   \\
    \midrule
     Beauty & 69.91\% & 15.40\% & 3.64\% &1.44\%\\
     Grocery \& Gourmet Food & 71.94\%& 21.22\%& 6.44\% & 2.75\%\\
    Health \& Personal Care& 69.04 \% &14.96\% & 3.76\% &1.62\%\\
  \bottomrule
\end{tabular}
\label{a}
\vspace{-5pt}
\end{table}

\begin{table}[t]
  \setlength{\tabcolsep}{4pt}
  \caption{Percentage of users located at different total consumption thresholds (as $P$ represents) in the training set of the datasets.}
  \label{tab:freq}
  \begin{tabular}{lllll}
    \toprule
    Dataset& $P\geq$50 &$P\geq$ 100 & $P\geq$200& $P\geq$ 400 \\
    \midrule
     Beauty & 71.23\% & 33.49\% & 10.07\% &2.22\%\\
     Grocery \& Gourmet Food & 90.87\%& 60.11\%& 20.35\% &5.62 \%\\
    Health \& Personal Care&  87.59\% &55.15\% & 20.59\% &5.78\%\\
  \bottomrule
\end{tabular}
\label{b}
\vspace{-5pt}
\end{table}

\begin{table}[t]
  \setlength{\tabcolsep}{4pt}
  \caption{Percentage of users located at different maximum price of purchase records thresholds (as $P$ represents) in the training set of the datasets.}
  \label{tab:freq}
  \begin{tabular}{lllll}
    \toprule
    Dataset& $P\geq$ 20&$P\geq$ 40 & $P\geq$80& $P\geq$ 160 \\
    \midrule
     Beauty &66.43 \% & 23.20\% & 6.14\% &1.47\%\\
     Grocery \& Gourmet Food & 92.90\%&44.38 \%& 2.40\% & 0.00\%\\
    Health \& Personal Care& 85.66 \% &46.39\% & 16.91\% &5.24\%\\
  \bottomrule
\end{tabular}
\label{c}
\vspace{-10pt}
\end{table}
% In fair machine learning research, the fairness concerns are usually from individual or group perspective.  Individual fairness requires that each similar individual should be treated similarly, while group fairness requires that the protected groups should be treated similarly to the advantaged group. The methods of distinguishing different groups in group fairness problem usually make use of intrinsic sensitive attributes \cite{chen2018fair, ekstrand2018all}. This paper considers to address the user-oriented group fairness in recommendation systems. However, the personal information which can be used as sensitive attributes are usually hard to obtain in commercial recommendation scenario, making it almost impossible to consider group fairness based on sensitive attributes of users in recommendations. In this paper, we consider to group users by their observable information, which is more practical to think about in commercial recommendation scenario. We explore three different methods to differentiate users into advantaged group and disadvantaged group to show their different level of visibility in recommender systems and consumption ability on e-commerce platform Amazon. Specifically, we can group users by: 1) the number of user-item interactions; 2) the sum consumption of users; 3) the maximum price of items which are bought by users. We claim that the bias existing between two groups can be learned by the learning algorithms and thus generate unfair recommendations between advantaged and disadvantaged groups. 
In this section, we aim to motivate fairness concerns by conducting data-driven observational analysis to show the unfair performance of current recommender systems.
More concretely, we access the imbalanced data distribution on three Amazon Review datasets: \textbf{Beauty}, \textbf{Grocery~\&~Gourmet Food} (\textbf{Grocery}), and \textbf{Health~\&~Personal Care} (\textbf{Health}), while the details of data are given in Table.\ref{data}. Furthermore, we show the recommendation performance ($\mathrm{F}_{1}@ 10$ and $\mathrm{NDCG}@ 10$) of several different kinds of traditional fairness-unaware recommendation algorithms on the three datasets to present their unfairness issue, including the shallow models biasedMF \cite{koren2009matrix} and PMF \cite{mnih2008probabilistic}, deep model NeuMF \cite{he2017neural}, and sequential recommendation algorithm STAMP \cite{liu2018stamp}.

In this section, we aim to show two facts. The first one is that the majority of users in these datasets only have limited interactions or consumption on the platform, and those users with much more interactions or much higher consumption only account for a small proportion of users. The second fact is that the average recommendation quality on this small group is significantly better than that on the remaining majority of users for all baselines.

\begin{figure}[t]
    \centering
        \pgfplotsset{label style={font=\normalsize},
                    tick label style={font=\normalsize}}
        \begin{subfigure}[b]{0.32\linewidth}   
            \centering 
            \includegraphics[scale=0.28]{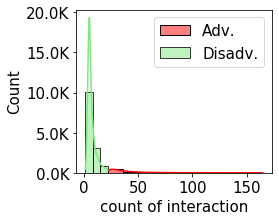}
            % \resizebox{\linewidth}{!}{
                
            % }
            \caption[]%
            {{\small interactions}}    
            \label{fig: ndcg_record}
        \end{subfigure}
        \hfill
        \begin{subfigure}[b]{0.32\linewidth}   
            \centering 
            \includegraphics[scale=0.28]{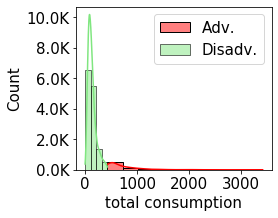}
            % \resizebox{\linewidth}{!}{
                
            % }
            \caption[]%
            {{\small total consumption}}    
            \label{fig:ndcg_sum_price}
        \end{subfigure}
        \hfill
        \begin{subfigure}[b]{0.32\linewidth}   
            \centering 
            \includegraphics[scale=0.28]{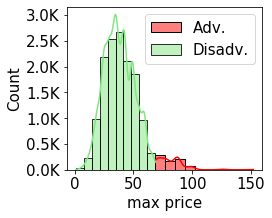}
            % \resizebox{\linewidth}{!}{
                
            % }
            \caption[]%
            {{\small max price}}    
            \label{fig:ndcg_max_price}
        \end{subfigure}
        % \vspace{-3pt}
        \caption[]{\small User distribution of the advantaged and disadvantaged groups under three grouping methods on the \textbf{Grocery} dataset.} 
        \label{distribution}
    \vspace{-10pt}
\end{figure}

Tables \ref{a},\ref{b},\ref{c} show the users' distribution in three datasets with different number of interactions or consumption.  We can see that users are concentrated in areas with less interaction or less consumption. Considering such imbalanced data distribution,  we select the top $5\%$ of users in the training dataset ranked by: 1) the number of interactions; 2) total consumption, i.e., the accumulated price of items bought by the user; and 3) the maximum price of items bought by the user, and label them as the advantaged group. The remaining users are labeled as the disadvantaged group. Intuitively, we present the distribution of users in advantaged and disadvantaged groups in \textbf{Grocery} as Figure.\ref{distribution}, which we can see clearly the difference between the two groups.

\begin{figure}[t]
    \centering
        \pgfplotsset{label style={font=\normalsize},
                    tick label style={font=\normalsize}}
        \begin{subfigure}[b]{0.32\linewidth}   
            \centering 
            \resizebox{\linewidth}{!}{
                \begin{tikzpicture}
                    \begin{axis}[
                        width  = 2*\textwidth,
                        major x tick style = transparent,
                        ybar=2*\pgflinewidth,
                        bar width=8pt,
                        ymajorgrids = true,
                        ylabel = {$\mathrm{F}_1@10$},
                        symbolic x coords={BiasedMF,PMF,NeuMF,STAMP},
                        xtick = data,
                        xticklabel style = {rotate=45,anchor=east},
                        scaled y ticks = false,
                        enlarge x limits=0.25,
                        ymin=0,
                        ymax=50,
                        legend image code/.code={%
                        \draw[#1, draw=none] (0cm,-0.1cm) rectangle (0.3cm,0.1cm);
                        },  
                        legend style={
                            draw=none, % ?
                            text depth=0pt,
                            at={(0.15,1.0)},
                            anchor=north west,
                            legend columns=-1,
                            font=\fontsize{10}{5}\selectfont,
                            % default spacing:
                            column sep=0.1cm,
                            % The text "Legend:"
                            % /tikz/column 2/.style={column sep=0pt,font=\bfseries},
                            %
                            % the space between legend image and text:
                            /tikz/every odd column/.append style={column sep=0cm},
                        }
                    ]
                        \addplot[style={celadon,fill=celadon,mark=none}]
                             coordinates {(BiasedMF,36.48) (PMF,36.37) (NeuMF,36.02) (STAMP,33.52)};
                        \addplot[style={red,fill=red!30,mark=none}]
                             coordinates {(BiasedMF,13.86) (PMF,13.70) (NeuMF,12.51) (STAMP,12.57)};
                        \addlegendentry{Adv.}
                        \addlegendentry{Disadv.}
                    \end{axis}
                \end{tikzpicture}
            }
            \caption[]%
            {{\small interactions}}
            % {{\small}}
            \label{fig:f1_record}
        \end{subfigure}
         \begin{subfigure}[b]{0.32\linewidth}   
            \centering 
          \resizebox{\linewidth}{!}{
                \begin{tikzpicture}
                    \begin{axis}[
                        width  = 2*\textwidth,
                        major x tick style = transparent,
                        ybar=2*\pgflinewidth,
                        bar width=8pt,
                        ymajorgrids = true,
                        ylabel = {$\mathrm{F}_1@10$},
                        symbolic x coords={BiasedMF,PMF,NeuMF,STAMP},
                        xtick = data,
                        xticklabel style = {rotate=45,anchor=east},
                        scaled y ticks = false,
                        enlarge x limits=0.25,
                        ymin=0,
                        ymax=45,
                        legend image code/.code={%
                        \draw[#1, draw=none] (0cm,-0.1cm) rectangle (0.3cm,0.1cm);
                        },  
                        legend style={
                            draw=none, % ?
                            text depth=0pt,
                            at={(0.15,1.0)},
                            anchor=north west,
                            legend columns=-1,
                            font=\fontsize{10}{5}\selectfont,
                            % default spacing:
                            column sep=0.1cm,
                            % The text "Legend:"
                            % /tikz/column 2/.style={column sep=0pt,font=\bfseries},
                            %
                            % the space between legend image and text:
                            /tikz/every odd column/.append style={column sep=0cm},
                        }
                    ]
                        \addplot[style={celadon,fill=celadon,mark=none}]
                             coordinates {(BiasedMF,33.85) (PMF,33.88) (NeuMF,33.16) (STAMP,30.94)};
                        \addplot[style={red,fill=red!30,mark=none}]
                             coordinates {(BiasedMF,14.15) (PMF,13.97) (NeuMF,12.82) (STAMP,12.85)};
                        \addlegendentry{Adv.}
                        \addlegendentry{Disadv.}
                    \end{axis}
                \end{tikzpicture}
            }
            \caption[]%
            {{\small total consumption}} 
            % {{\small}}    
            \label{fig}
        \end{subfigure}
         \begin{subfigure}[b]{0.32\linewidth}   
            \centering 
            \resizebox{\linewidth}{!}{
                \begin{tikzpicture}
                    \begin{axis}[
                        width  = 2*\textwidth,
                        major x tick style = transparent,
                        ybar=2*\pgflinewidth,
                        bar width=8pt,
                        ymajorgrids = true,
                        ylabel = {$\mathrm{F}_1@10$},
                        symbolic x coords={BiasedMF,PMF,NeuMF,STAMP},
                        xtick = data,
                        xticklabel style = {rotate=45,anchor=east},
                        scaled y ticks = false,
                        enlarge x limits=0.25,
                        ymin=0,
                        ymax=33,
                        legend image code/.code={%
                        \draw[#1, draw=none] (0cm,-0.1cm) rectangle (0.3cm,0.1cm);
                        },  
                        legend style={
                            draw=none, % ?
                            text depth=0pt,
                            at={(0.15,1.0)},
                            anchor=north west,
                            legend columns=-1,
                            font=\fontsize{10}{5}\selectfont,
                            % default spacing:
                            column sep=0.1cm,
                            % The text "Legend:"
                            % /tikz/column 2/.style={column sep=0pt,font=\bfseries},
                            %
                            % the space between legend image and text:
                            /tikz/every odd column/.append style={column sep=0cm},
                        }
                    ]
                        \addplot[style={celadon,fill=celadon,mark=none}]
                             coordinates {(BiasedMF,24.29) (PMF,24.09) (NeuMF,22.84) (STAMP,21.75)};
                        \addplot[style={red,fill=red!30,mark=none}]
                             coordinates {(BiasedMF,15.25) (PMF,15.08)(NeuMF,13.98) (STAMP,13.96)};
                        \addlegendentry{Adv.}
                        \addlegendentry{Disadv.}
                    \end{axis}
                \end{tikzpicture}
            }
            \caption[]%
            {{\small max price}} 
            % {{\small}}  
            \label{fig:f1_max_price}
        \end{subfigure}
        \vspace{-3pt}
        \caption[]{\small The difference between advantaged group and disadvantaged group on $\mathrm{F}_{1}@ 10$ on the \textbf{Grocery} dataset.}
        \label{Ff1}
        \vspace{-15pt}
\end{figure}
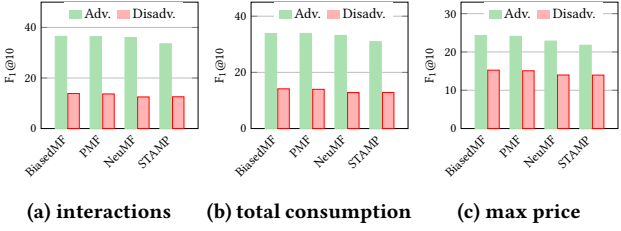

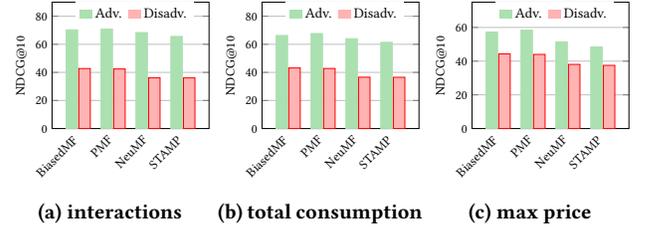
\begin{figure}[t]
    \centering
        \pgfplotsset{label style={font=\normalsize},
                    tick label style={font=\normalsize}}
        \begin{subfigure}[b]{0.32\linewidth}   
            \centering 
            \resizebox{\linewidth}{!}{
                \begin{tikzpicture}
                    \begin{axis}[
                        width  = 2*\textwidth,
                        major x tick style = transparent,
                        ybar=2*\pgflinewidth,
                        bar width=8pt,
                        ymajorgrids = true,
                        ylabel = {NDCG@10},
                        symbolic x coords={BiasedMF,PMF,NeuMF,STAMP},
                        xtick = data,
                        xticklabel style = {rotate=45,anchor=east},
                        scaled y ticks = false,
                        enlarge x limits=0.25,
                        ymin=0,
                        ymax=90,
                        legend image code/.code={%
                        \draw[#1, draw=none] (0cm,-0.1cm) rectangle (0.3cm,0.1cm);
                        },  
                        legend style={
                            draw=none, % ?
                            text depth=0pt,
                            at={(0.15,1.0)},
                            anchor=north west,
                            legend columns=-1,
                            font=\fontsize{10}{5}\selectfont,
                            % default spacing:
                            column sep=0.1cm,
                            % The text "Legend:"
                            % /tikz/column 2/.style={column sep=0pt,font=\bfseries},
                            %
                            % the space between legend image and text:
                            /tikz/every odd column/.append style={column sep=0cm},
                        }
                    ]
                        \addplot[style={celadon,fill=celadon,mark=none}]
                             coordinates {(BiasedMF,70.29) (PMF,70.80) (NeuMF,68.34) (STAMP,65.61)};
                        \addplot[style={red,fill=red!30,mark=none}]
                             coordinates {(BiasedMF,42.74) (PMF,42.44) (NeuMF,36.13) (STAMP,36.08)};
                        \addlegendentry{Adv.}
                        \addlegendentry{Disadv.}
                    \end{axis}
                \end{tikzpicture}
            }
            \caption[]%
            {{\small interactions}}    
            \label{fig: ndcg_record}
        \end{subfigure}
        \begin{subfigure}[b]{0.32\linewidth}   
            \centering 
            \resizebox{\linewidth}{!}{
                \begin{tikzpicture}
                    \begin{axis}[
                        width  = 2*\textwidth,
                        major x tick style = transparent,
                        ybar=2*\pgflinewidth,
                        bar width=8pt,
                        ymajorgrids = true,
                        ylabel = {NDCG@10},
                        symbolic x coords={BiasedMF,PMF,NeuMF,STAMP},
                        xtick = data,
                        xticklabel style = {rotate=45,anchor=east},
                        scaled y ticks = false,
                        enlarge x limits=0.25,
                        ymin=0,
                        ymax=90,
                        legend image code/.code={%
                        \draw[#1, draw=none] (0cm,-0.1cm) rectangle (0.3cm,0.1cm);
                        },  
                        legend style={
                            draw=none, % ?
                            text depth=0pt,
                            at={(0.15,1.00)},
                            anchor=north west,
                            legend columns=-1,
                            font=\fontsize{10}{5}\selectfont,
                            % default spacing:
                            column sep=0.1cm,
                            % The text "Legend:"
                            % /tikz/column 2/.style={column sep=0pt,font=\bfseries},
                            %
                            % the space between legend image and text:
                            /tikz/every odd column/.append style={column sep=0cm},
                        }
                    ]
                        \addplot[style={celadon,fill=celadon,mark=none}]
                             coordinates {(BiasedMF,66.29)(PMF,67.67) (NeuMF,64.02) (STAMP,61.41)};
                        \addplot[style={red,fill=red!30,mark=none}]
                             coordinates {(BiasedMF,43.16) (PMF,42.78)(NeuMF,36.59) (STAMP,36.52)};
                        \addlegendentry{Adv.}
                        \addlegendentry{Disadv.}
                    \end{axis}
                \end{tikzpicture}
            }
            \caption[]%
            {{\small total consumption}}    
            \label{fig:ndcg_sum_price}
        \end{subfigure}
        \begin{subfigure}[b]{0.32\linewidth}   
            \centering 
            \resizebox{\linewidth}{!}{
                \begin{tikzpicture}
                    \begin{axis}[
                        width  = 2*\textwidth,
                        major x tick style = transparent,
                        ybar=2*\pgflinewidth,
                        bar width=8pt,
                        ymajorgrids = true,
                        ylabel = {NDCG@10},
                        symbolic x coords={BiasedMF,PMF,NeuMF,STAMP},
                        xtick = data,
                        xticklabel style = {rotate=45,anchor=east},
                        scaled y ticks = false,
                        enlarge x limits=0.25,
                        ymin=0,
                        ymax=75,
                        legend image code/.code={%
                        \draw[#1, draw=none] (0cm,-0.1cm) rectangle (0.3cm,0.1cm);
                        },  
                        legend style={
                            draw=none, % ?
                            text depth=0pt,
                            at={(0.15,1.0)},
                            anchor=north west,
                            legend columns=-1,
                            font=\fontsize{10}{5}\selectfont,
                            % default spacing:
                            column sep=0.1cm,
                            % The text "Legend:"
                            % /tikz/column 2/.style={column sep=0pt,font=\bfseries},
                            %
                            % the space between legend image and text:
                            /tikz/every odd column/.append style={column sep=0cm},
                        }
                    ]
                        \addplot[style={celadon,fill=celadon,mark=none}]
                             coordinates {(BiasedMF,57.29)(PMF,58.39) (NeuMF,51.48) (STAMP,48.39)};
                        \addplot[style={red,fill=red!30,mark=none}]
                             coordinates {(BiasedMF,44.30)(PMF,43.99) (NeuMF,38.07) (STAMP,37.51)};
                        \addlegendentry{Adv.}
                        \addlegendentry{Disadv.}
                    \end{axis}
                \end{tikzpicture}
            }
            \caption[]%
            {{\small max price}}    
            \label{N}
        \end{subfigure}
        \vspace{-3pt}
        \caption[]{\small The difference between advantaged group and disadvantaged group on $\mathrm{NDCG}@ 10$ on the \textbf{Grocery} dataset.} 
        \label{Fndcg}
    \vspace{-8pt}
\end{figure}

Next, we test the recommendation quality of the four baselines on the three datasets. Here we show the recommendation quality of these fairness-unaware recommendation algorithms on \textbf{Grocery} in Figure \ref{Ff1} and \ref{Fndcg}.  Similar trends are observed for the other two datasets as well. The details of the experiment results are given in Table \ref{record}, Table \ref{sum}, and Table \ref{max} in later sections. We can see that although the advantaged user group only accounts for a very small proportion of users $(5\%)$, they enjoy much higher recommendation quality than those disadvantaged users. This reflects the majority of users are easily disregarded by commercial recommendation engines, which gives rise to unfair recommendations, as well as results in degradation of the overall performance. Therefore, it is important to devise techniques to better serve such users with higher quality recommendations to encourage them to make further interactions with the system, and also to improve the overall recommendation quality since the disadvantaged users are the vast majority.

\section{The Framework}
\label{sec:framework}

In order to address the unfairness concerns presented in Section \ref{sec:motivation}, we provide a framework in this section to generate fair recommendation lists for different user groups, which also has the ability to improve the overall performance through providing more satisfying recommendations to the majority disadvantaged users. We first give the definition of user-oriented group fairness in recommendation systems, and then provide our re-ranking method to formalize the fair recommendation problem under fairness constraints.

In the problem of recommendation, suppose there are user set $\left\{u_{1}, u_{2}, \cdots, u_{n}\right\}\in \mathcal{U}$ and item set $\left\{v_{1}, v_{2}, \cdots, v_{m}\right\}\in \mathcal{V}$, where $n=|\mathcal{U}|$, $m=|\mathcal{V}|$. Given a recommender system, each user $u_{i}$ will have a top-$N$  recommendation list $\left\{v_{1}, v_{2}, \cdots, v_{N} | u_{i}\right\} .$ As we analyzed in the previous section, recommender systems without considering the user-oriented fairness will generate unfair recommendation lists to different groups of users. To address this issue, we provide a re-ranking framework to generate fair recommendations based on the recommendation lists produced by traditional fairness-unaware recommender systems.

We define binary matrix $\mathbf{W}=\left[\mathbf{W}_{i j}\right]_{n \times N}$ to denote whether an item $j$ is recommended to a user $i$ in fair recommendation lists, where $\mathbf{W}_{i j} \in\{0,1\}$,  and $\{1 \leq i \leq n\}$ and $\{1 \leq j \leq N\}$ index users and items respectively. Specifically, if item $j$ is recommended to the user $i$, then we have $\mathbf{W}_{i j} = 1$, else $\mathbf{W}_{i j} = 0$.  We use $\mathbf{W}_{\mathrm{i}}=\left[\mathbf{W}_{i 1}, \mathbf{W}_{i 2}, \cdots ,\mathbf{W}_{i N}\right]^{\mathrm{T}}$ to represent the new Top-$K$ recommendation list of user $i$, where $\sum_{j=1}^{N} \mathbf{W}_{i j}=K, K \leq N$. Next we define the user-oriented group fairness in recommender systems following these notations.

\subsection{User-oriented Group Fairness}
Group fairness requires that the protected groups should be treated similarly to the advantaged group \cite{pedreschi2009measuring}. The group of users can be divided under different requirements for different tasks. In this paper, we consider grouping users as $Z_{1}$ and $Z_{2}$ so that $Z_{1} \cap Z_{2}=\varnothing$ in accordance with their different activity level. The notation $\mathcal{M}$ is a metric that can evaluate the recommendation quality such as $\text{NDCG} @ K$ or $\mathrm{F}_{1}$ score, and thus we use $\mathcal{M}\left(\mathrm{W}_{\mathrm{i}}\right)$ to represent the recommendation quality for user $i$. 

The user-oriented group fairness in recommendation is defined as follows:

%%%%%%%%%%%%%%%%%%%%%%%%
% set equation space 
\setlength{\belowdisplayskip}{0pt} \setlength{\belowdisplayshortskip}{0pt}
\setlength{\abovedisplayskip}{-5pt} \setlength{\abovedisplayshortskip}{-10pt}
%%%%%%%%%%%%%%%%%%%%%%%%

\begin{definition}
  User-oriented Group Fairness (UGF)
\end{definition}
\begin{equation}
\mathbb{E}[\mathcal{M}
(\mathbf{W})| Z=Z_{1}]=\mathbb{E}[\mathcal{M}
(\mathbf{W})| Z=Z_{2}]
\label{fa}
\vspace{5pt}
\end{equation}

% The item-oriented group fairness is defined as the following:

% \begin{definition}
%   Item-oriented Group Fairness (IGF)
% \end{definition}

% \begin{equation}
% \mathbb{E}[\mathcal{H}
% (\mathbf{Q})| W=I_{1}]=\mathbb{E}[\mathcal{H}
% (\mathbf{Q})| W=I_{2}]
% \end{equation}

% \begin{equation}
%     UGF\left(G_{1}, G_{2}, \mathbf{Q}\right)=\left|\frac{1}{\left|G_{1}\right|} \sum_{i \in G_{1}} \mathcal{F}\left(\mathbf{Q}_{\mathbf{i}}\right)-\frac{1}{\left|G_{2}\right|} \sum_{i \in G_{2}} \mathcal{F}\left(\mathbf{Q}_{\mathbf{i}}\right)\right|
% \end{equation}

It requires that a fair recommendation algorithm should offer the same recommendation quality for different group of users. Furthermore, we use the difference of average recommendation performance between two groups to measure the user-oriented group unfairness of a recommendation algorithm. We define $\varepsilon$-fairness recommendation algorithm as:

\begin{definition}
  ($\varepsilon$-fairness) A recommendation algorithm satisfies $\varepsilon$-fairness if:
  \vspace{-15pt}
\end{definition}
\begin{equation}
\small
    UGF\left(Z_{1}, Z_{2}, \mathbf{W}\right)=\left|\frac{1}{\left|Z_{1}\right|} \sum_{i \in Z_{1}} \mathcal{M}\left(\mathbf{W}_{\mathbf{i}}\right)-\frac{1}{\left|Z_{2}\right|} \sum_{i \in Z_{2}} \mathcal{M}\left(\mathbf{W}_{\mathbf{i}}\right)\right| \leq \varepsilon.
\label{fair}
\end{equation}

In this formulation, $\epsilon$ represents the strictness of fairness requirements. It also trades off fairness and the recommendation quality of the advantaged group, so that if $\epsilon$ approaches to zero, the resulting recommendation will be fairer but will potentially suffer from a huge sacrifice in the recommendation performance of the advantaged group.

% \begin{equation}
% I G F\left(I_{1}, I_{2}, \mathbf{Q}\right)=\left|\frac{1}{|U|} \sum_{j \in I_{1}} \sum_{i=1}^{m} \mathbf{Q}_{\mathrm{ij}}-\frac{1}{|U|} \sum_{j \in I_{2}} \sum_{i=1}^{m} \mathbf{Q}_{\mathrm{ij}}\right| \leq \varepsilon_{2}
% \end{equation}

\subsection{Fairness-aware Algorithm}

In this part, we provide a framework which can generate fairness-aware recommendation lists based on a fairness-constrained re-ranking method.

Given a traditional recommender system, each user $u_{i}$ is recommended with a top-$N$  list $\left\{v_{1}, v_{2}, \cdots, v_{N} | u_{i}\right\}$, each user-item pair is associated with a score $\mathrm{S}_{i,j}$ which represents the preference of user $i$ in terms of item $j$. Here we follow the preference scores calculated by the base recommendation systems. We use these system generated top-$N$ ranking lists as the baseline, and apply re-ranking algorithm to maximize the sum of preference scores under the fairness constraint to generate new fair top-$K$ recommendation lists, where $K \leq N$. Therefore, we can formulate the optimization procedure of the fairness-aware recommendation problem as follows:

%%%%%%%%%%%%%%%%%%%%%%%%
% set equation space 
\setlength{\belowdisplayskip}{-10pt} \setlength{\belowdisplayshortskip}{-10pt}
\setlength{\abovedisplayskip}{-5pt} \setlength{\abovedisplayshortskip}{-5pt}
%%%%%%%%%%%%%%%%%%%%%%%%

\begin{equation}
\begin{split}
 \max _{\mathbf{W}_{i j}} & \quad 
% \quad \sum_{i=1}^{n} \mathcal{L}\left(\mathbf{W}_{\mathrm{i}}\right)= 
\sum_{i=1}^{n}\sum_{j=1}^{N} \mathbf{W}_{i j} \mathrm{S}_{i,j} \\
\text { s.t. } & \quad \operatorname{UGF}\left(Z_{1}, Z_{2}, \mathbf{W}\right)<\varepsilon\\
&\quad \sum_{j=1}^{N} \mathbf{W}_{i j}=K, \mathbf{W}_{i j} \in\{0,1\} \\
\\
% & \quad \operatorname{IGF}\left(I_{1}, I_{2}, \mathrm{Q}\right)<\varepsilon_{2}
\end{split}
\end{equation}

This objective function can be interpreted as that selecting exactly $K$ items out of the baseline top-$N$ list of each user so that the objective function could be maximized. Meanwhile, these selected items need to make the new top-$K$ recommendation list satisfy the constraint defined in Definition~\ref{fair}.  The optimization problem here can be solved as a 0–1 integer programming problem. We can find feasible solutions to this NP-complete problem through fast heuristics\footnote{We use gurobi solver in our experiment. https://www.gurobi.com}. Although such methods may not converge to the global optimal solution, our experiments show that we can still obtain satisfactory results in this way. After obtaining the item set which is recommended under fairness constrained, we rank the $K$ items by their original preference score $\mathrm{S}_{i,j}$ to construct the final recommendation list.

\section{Experiments}
\label{sec:experiments}

In this section, we first briefly describe the datasets, baselines and experimental setup used for experiments. All source code and dataset of this project has been released publicly\footnote{Source code available at \url{https://github.com/rutgerswiselab/user-fairness}}. Then we evaluate our proposed fair re-ranking algorithm on top of the baselines to show its desirable performance on both of the fairness metrics and the recommendation performance.

\subsection{Experimental Settings}
\begin{table}
  \setlength{\tabcolsep}{4pt}
  \caption{Statistics of the datasets}
  \label{tab:freq}
  \begin{tabular}{lllll}
    \toprule
    Dataset& \#Action & \#User & \#Item & Sparsity  \\
    \midrule
     Beauty & 198,502& 22,363& 12,101&99.93\%\\
     Grocery \& Gourmet Food &151,254 &14,681 &8,713 &99.88\%\\
    Health \& Personal Care&346,355  &38,609 &18,534 &99.95\%\\
  \bottomrule
 \label{data}
\end{tabular}
\vspace{-20pt}
\end{table}

\subsubsection{\textbf{Dataset}}
Our experiments are performed on publicly available Amazon 5-core datasets\footnote{http://jmcauley.ucsd.edu/data/amazon/}, which include user, item, rating information spanning from May 1996 to July 2014 without duplicated interactions. It covers 24 different categories and we take three datasets \textbf{Beauty}, \textbf{Grocery~\&~Gourmet Food} (\textbf{Grocery}), and \textbf{Health~\&~Personal Care} (\textbf{Health}) to model training and evaluation. The statistics of the datasets are summarized in Table~\ref{tab:freq}. In our experiments, we split each dataset into train ($80\%$), validation ($10\%$) and test sets ($10\%$) and all the baseline models share these datasets for training and evaluation. 

For the re-ranking experiment, as stated in Section~\ref{MFC}, we select the top 5$\%$ users under each grouping type from the training set as the advantaged group and the rest as the disadvantaged group. Then we split the test set based on the two user groups and test the results, respectively. 

\begin{table*}[t]
% \caption{Ranking performance under different logical structures. GRU4Rec is the best baseline copied from Table~\ref{tb:results}. "*" indicates the significance at the level of 0.05}\label{tb:logic_structure}
\caption{The recommendation performance of overall, advantaged, and disadvantaged users of our re-ranking method and corresponding baselines on three Amazon datasets, with the type of grouping users by their number of interactions. The results are reported in percentage (\%). All re-ranking results here are obtained under the fairness constraint on $\mathrm{F}_{1}$. The evaluation metrics here are calculated based on the top-10 predictions in the test set. Our best results are highlighted in bold.}\label{record}
% \vspace{-5pt}
\centering\begin{tabular}{lccc
ccccccccccc}
\toprule
\multirow{2}{*}{}& & &\multicolumn{4}{c}{\textbf{Beauty}} & \multicolumn{4}{c}{\textbf{Grocery}} & \multicolumn{4}{c}{\textbf{Health}} \\ 
\cmidrule(lr){4-7}
\cmidrule(lr){8-11}
\cmidrule(lr){12-15}
& & & Overall &Adv. &Disadv. &UGF &Overall &Adv. &Disadv. & UGF &Overall &Adv. &Disadv. &UGF \\
\midrule
%  RNNModel &0.3593 &0.4211 &0.5149 &0.7059 &0.3697 &0.4119 &0.5035 &0.6338 & 0.3162 & 0.3555 & 0.4289 & 0.5560\\
%  CMP-I &0.3537 & 0.4093 & 0.5202 & 0.6921 &0.3952 &0.4364 &0.5265 &0.6539 &0.3205 &0.3578 &0.4282 &0.5434\\
\multirow{4}{*}{BiasedMF}&
\multirow{2}{*}{F1} &
Orig. &14.27 &30.68 & 12.77&17.91&15.78&36.48&13.86&22.62&13.92&33.06&12.13&20.93\\
&&Fair &\textbf{15.06}&19.18&\textbf{14.68}&\textbf{4.50}&\textbf{16.13}&21.37&\textbf{15.65}&\textbf{5.72}&\textbf{14.54}&19.30&\textbf{14.10}&\textbf{5.20}\\
 \cline{2-15}
&\multirow{2}{*}{NDCG}&
Orig. &43.25&67.79&41.00&26.79&45.08&70.29&42.74&27.55&41.37&66.55&39.01&27.54\\
&&Fair &\textbf{43.97}&52.51&\textbf{43.19}&\textbf{9.32}&\textbf{45.75}&55.74&\textbf{44.83}&\textbf{10.91}&\textbf{42.31}&52.26&\textbf{41.37}&\textbf{10.89}\\ 
\hline
\multirow{4}{*}{PMF}&
\multirow{2}{*}{F1} &
Orig. &13.72&30.87&12.15&18.72&15.62&36.37&13.70&22.67&11.34&22.46&10.30&12.16\\
&&Fair &\textbf{14.56}&18.86&\textbf{14.16}&\textbf{4.70}&\textbf{16.06}&21.28&\textbf{15.58}&\textbf{5.70}&\textbf{11.42}&14.16&\textbf{11.16}&\textbf{3.00}\\
 \cline{2-15}
&\multirow{2}{*}{NDCG}&
Orig. &41.06&66.74&38.72&28.02&44.85&70.80&42.44&28.36&36.23&52.43&34.72&17.71\\
&&Fair &\textbf{41.97}&51.41&\textbf{41.10}&\textbf{10.31}&\textbf{45.74}&57.23&\textbf{44.67}&\textbf{12.56}&\textbf{36.75}&46.39&\textbf{35.85}&\textbf{10.54}\\ 
\hline
\multirow{4}{*}{NeuMF}&
\multirow{2}{*}{F1} &
Orig. &12.44&29.10&10.91&18.19&14.51&36.02&12.51&23.51&12.20&31.84&10.36&21.48\\
&&Fair &\textbf{13.81}&17.93&\textbf{13.43}&\textbf{3.50}&\textbf{15.48}&20.81&\textbf{14.99}&\textbf{5.82}&\textbf{12.96}&17.89&\textbf{12.49}&\textbf{5.40}\\
 \cline{2-15}
&\multirow{2}{*}{NDCG}&
Orig. &35.13&61.76&32.69&29.07&38.86&68.34&36.13&32.21&33.55&61.72&30.91&30.81\\
&&Fair &\textbf{36.30}&40.89&\textbf{35.88}&\textbf{5.01}&\textbf{40.09}&50.02&\textbf{39.17}&\textbf{10.85}&\textbf{34.30}&42.57&\textbf{33.53}&\textbf{9.04}\\ 
\hline
\multirow{4}{*}{STAMP}&
\multirow{2}{*}{F1} &
Orig. &12.76&27.68&11.38&16.30&14.35&33.52&12.57&20.95&13.15&30.76&11.50&19.26\\
&&Fair &12.76&20.27&\textbf{12.07}&\textbf{8.20}&\textbf{14.47}&19.23&\textbf{14.03}&\textbf{5.20}&13.15&17.55&\textbf{12.74}&\textbf{4.81}\\
 \cline{2-15}
&\multirow{2}{*}{NDCG}&
Orig. &35.54&58.32&33.45&24.87&38.58&65.61&36.08&29.53&36.53&61.06&34.23&26.83\\
&&Fair &\textbf{35.71}&51.02&\textbf{34.31}&\textbf{16.71}&\textbf{39.16}&52.42&\textbf{37.93}&\textbf{14.49}&\textbf{36.69}&46.91&\textbf{35.73}&\textbf{11.18}\\ 
 \bottomrule
%  NCR-Reg &0.00 &0.00 &0.00 &0.00 &0.00 &0.00 \\ 
%  \hline
\end{tabular}
\vspace{-9pt}
\end{table*}

\subsubsection{\textbf{Baselines}} We take both shallow and deep recommendation models as baselines as suggested in \cite{dacrema2019we}.
We compare with two traditional shallow methods (Biased-MF and PMF), one deep model (NeuMF), as well as one sequential model (STAMP). The introduction of baselines are as the following:
\begin{itemize}
    \item \textbf{Biased-MF} \cite{koren2009matrix}: This matrix factorization algorithm takes user, item and global bias terms into consideration. 
    \item \textbf{PMF} \cite{mnih2008probabilistic}: This is a probabilistic matrix factorization algorithm by adding Gaussian prior into the user and item latent factors distribution. 
    % \item \textbf{DMF} \cite{xue2017deep}: Deep Matrix Factorization is a deep model for recommendation, which uses multiple non-linear layers to process the raw user-item interaction matrix. 
    \item \textbf{NeuMF} \cite{he2017neural}: This algorithm applies deep neural network with non-linear activation functions to train a user and item matching function.
    % It contains two parts, Generalized Matrix Factorization (GMF) and multi-layer perceptron (MLP). We implement their model by combining both modules together for training.
    % \item \textbf{GRU4Rec} \cite{hidasi2016session}: This is a well-known session-based recommendation model, which captures the sequential dependencies in user's historical interactions through recurrent neural network (RNN).
    \item \textbf{STAMP} \cite{liu2018stamp}: A session-based recommendation model based on attention mechanism, which can capture user's long-term and short-term preferences.
\end{itemize}

We set the embedding size for users and items to 64 for all the models. For NeuMF, we set the size of multi-layer perceptron (MLP) with 32, 16, 8 as suggested in the paper. The final output layer has only one layer with a dimension of 64. For STAMP, we set the maximum user history length to 30. We apply rectified linear unit (ReLU) non-linear activation function between layers. 

We apply Bayesian Personalized Ranking (BPR)~\cite{rendle2012bpr} loss for all the baseline models. For each user-item pair in the training dataset, we randomly sample one item that the user has never interacted with as the negative sample in one training epoch. We carefully select the hyper-parameters to tune the models to reach their best performance. The learning rate for training is 0.001, $\ell2$-regularization coefficient is 0.00001 for all the datasets. The best models are selected based on the performance on the validation set within 100 epochs. 
\vspace{-5pt}
\subsubsection{\textbf{Evaluation}}We use standard metrics $\mathrm{F}_{1}@ 10$ score and  Normalized Discounted accumulated Gain at rank $10$ ($\mathrm{NDCG}@ 10$) to evaluate the top-$10$ recommendation quality. The metric $\mathcal{M}$ given in Definition~\ref{fair} is F$_1@10$ in all our experiments. For efficiency consideration, we use sampled negative interactions for evaluation instead of computing the user-item pairs scores for each user over the entire item space \cite{zhao2020revisiting}. For each user, we randomly generate 100 negative samples, which the user has never interacted with, together with the positive samples in the validation or test set to form the user's candidates list. Then we compute the metric scores over this candidates list to evaluate the models top-$K$ ranking performance. The result of all metrics in our experiments are averaged over all users.

\begin{table*}[t]
% \caption{Ranking performance under different logical structures. GRU4Rec is the best baseline copied from Table~\ref{tb:results}. "*" indicates the significance at the level of 0.05}\label{tb:logic_structure}
\caption{The recommendation performance of overall, advantaged, and disadvantaged users of our re-ranking method and corresponding baselines on three Amazon datasets, with the method of grouping users by their total consumption. The results are reported in percentage (\%). All re-ranking results here are obtained under the fairness constraint on $\mathrm{F}_{1}$. The evaluation metrics here are calculated based on the top-10 predictions in the test set. Our best results are highlighted in bold.}\label{sum}
% \vspace{-5pt}
\centering\begin{tabular}{lccc
ccccccccccc}
\toprule
\multirow{2}{*}{}& & &\multicolumn{4}{c}{\textbf{Beauty}} & \multicolumn{4}{c}{\textbf{Grocery}} & \multicolumn{4}{c}{\textbf{Health}} \\ 
\cmidrule(lr){4-7}
\cmidrule(lr){8-11}
\cmidrule(lr){12-15}
& & & Overall &Adv. &Disadv. &UGF &Overall &Adv. &Disadv. & UGF &Overall &Adv. &Disadv. &UGF \\
\midrule
%  RNNModel &0.3593 &0.4211 &0.5149 &0.7059 &0.3697 &0.4119 &0.5035 &0.6338 & 0.3162 & 0.3555 & 0.4289 & 0.5560\\
%  CMP-I &0.3537 & 0.4093 & 0.5202 & 0.6921 &0.3952 &0.4364 &0.5265 &0.6539 &0.3205 &0.3578 &0.4282 &0.5434\\
\multirow{4}{*}{BiasedMF}&
\multirow{2}{*}{F1} &
Orig. &14.27&29.76&13.01&16.75&15.78&33.85&14.15&19.70&13.92&33.04&12.31&20.73\\
&&Fair &\textbf{15.05}&16.91&\textbf{14.90}&\textbf{2.01}&\textbf{16.14}&17.99&\textbf{15.97}&\textbf{2.02}&\textbf{14.50}&19.11&\textbf{14.11}&\textbf{5.00}\\
 \cline{2-15}
&\multirow{2}{*}{NDCG}&
Orig. &43.25&65.94&41.40&24.54&45.08&66.29&43.16&23.13&41.37&67.72&39.15&28.57\\
&&Fair &\textbf{43.82}&47.07&\textbf{43.55}&\textbf{3.52}&\textbf{45.49}&47.97&\textbf{45.27}&\textbf{2.70}&\textbf{42.15}&52.20&\textbf{41.31}&\textbf{10.89}\\ 
\hline
\multirow{4}{*}{PMF}&
\multirow{2}{*}{F1} &
Orig. &13.72&29.74&12.42&17.32&15.62&33.88&13.97&19.91&11.34&22.79&10.38&12.41\\
&&Fair &\textbf{14.56}&18,54&\textbf{14.23}&\textbf{4.31}&\textbf{16.00}&20.59&\textbf{15.58}&\textbf{5.01}&\textbf{11.42}&14.28&\textbf{11.17}&\textbf{3.11}\\
 \cline{2-15}
&\multirow{2}{*}{NDCG}&
Orig. &41.06&64.57&39.15&25.42&44.85&67.67&42.78&24.89&36.23&54.80&34.67&20.13\\
&&Fair &\textbf{41.98}&50.35&\textbf{41.29}&\textbf{9.06}&\textbf{45.52}&54.40&\textbf{44.71}&\textbf{9.69}&\textbf{36.65}&47.93&\textbf{35.70}&\textbf{12.23}\\ 
\hline
\multirow{4}{*}{NeuMF}&
\multirow{2}{*}{F1} &
Orig. &12.44&28.28&11.15&17.13&14.51&33.16&12.82&20.34&12.20&32.10&10.53&21.57\\
&&Fair &\textbf{14.04}&15.91&\textbf{13.89}&\textbf{2.02}&\textbf{15.49}&17.34&\textbf{15.32}&\textbf{2.02}&\textbf{13.06}&14.91&\textbf{12.91}&\textbf{2.00}\\
 \cline{2-15}
&\multirow{2}{*}{NDCG}&
Orig. &35.13&60.00&33.10&26.90&38.86&64.02&36.59&27.43&33.55&61.74&31.17&30.57\\
&&Fair &\textbf{36.71}&39.11&\textbf{36.51}&\textbf{2.60}&\textbf{39.84}&41.55&\textbf{39.69}&\textbf{1.86}&\textbf{34.39}&38.60&\textbf{34.04}&\textbf{4.56}\\ 
\hline
\multirow{4}{*}{STAMP}&
\multirow{2}{*}{F1} &
Orig. &12.75&26.71&11.61&15.10&14.35&30.94&12.85&18.09&13.15&30.85&11.66&19.19\\
&&Fair &\textbf{12.81}&16.32&\textbf{12.53}&\textbf{3.79}&\textbf{14.45}&18.58&\textbf{14.08}&\textbf{4.50}&13.15&17.58&\textbf{12.78}&\textbf{4.80}\\
 \cline{2-15}
&\multirow{2}{*}{NDCG}&
Orig. &35.54&57.38&33.75&23.63&38.59&61.41&36.52&24.89&36.53&61.34&34.44&26.90\\
&&Fair &\textbf{35.70}&46.08&\textbf{34.85}&\textbf{11.23}&\textbf{38.86}&47.77&\textbf{38.06}&\textbf{9.71}&\textbf{36.66}&47.31&\textbf{35.76}&\textbf{11.55}\\ 
 \bottomrule
\end{tabular}
\vspace{-5pt}
\end{table*}

\begin{table*}[t]
% \caption{Ranking performance under different logical structures. GRU4Rec is the best baseline copied from Table~\ref{tb:results}. "*" indicates the significance at the level of 0.05}\label{tb:logic_structure}
\caption{The recommendation performance of overall, advantaged, and disadvantaged users of our re-ranking method and corresponding baselines on three Amazon datasets, with the method of grouping users by the maximum price of items they bought. The results are reported in percentage (\%). All re-ranking results here are obtained under the fairness constraint on $\mathrm{F}_{1}$. The evaluation metrics are calculated based on the top-10 predictions in the test set. Our best results are highlighted in bold.}\label{max}
% \vspace{-5pt}
\centering\begin{tabular}{lccc
ccccccccccc}
\toprule
\multirow{2}{*}{}& & &\multicolumn{4}{c}{\textbf{Beauty}} & \multicolumn{4}{c}{\textbf{Grocery}} & \multicolumn{4}{c}{\textbf{Health}} \\ 
\cmidrule(lr){4-7}
\cmidrule(lr){8-11}
\cmidrule(lr){12-15}
& & & Overall &Adv. &Disadv. &UGF &Overall &Adv. &Disadv. & UGF &Overall &Adv. &Disadv. &UGF \\
\midrule
%  RNNModel &0.3593 &0.4211 &0.5149 &0.7059 &0.3697 &0.4119 &0.5035 &0.6338 & 0.3162 & 0.3555 & 0.4289 & 0.5560\\
%  CMP-I &0.3537 & 0.4093 & 0.5202 & 0.6921 &0.3952 &0.4364 &0.5265 &0.6539 &0.3205 &0.3578 &0.4282 &0.5434\\
\multirow{4}{*}{BiasedMF}&
\multirow{2}{*}{F1} &
Orig. &14.27 &21.31 &13.86 &7.45 &15.78 &24.29&15.25&9.04&13.92&22.21&13.41&8.80\\
&&Fair &\textbf{14.57}&15.51&\textbf{14.51}&\textbf{1.00}&\textbf{15.84}&20.55&\textbf{15.55}&\textbf{5.00}&\textbf{14.15}&15.09&\textbf{14.09}&\textbf{1.00}\\
 \cline{2-15}
&\multirow{2}{*}{NDCG}&
Orig. &43.25&53.59&42.63&10.96&45.08&57.29&44.30&12.99&41.37&53.16&40.64&12.52\\
&&Fair &\textbf{43.44}&44.56&\textbf{43.37}&\textbf{1.19}&\textbf{45.09}&52.34&\textbf{44.63}&\textbf{7.71}&\textbf{41.44}&41.81&\textbf{41.41}&\textbf{0.40}\\ 
\hline
\multirow{4}{*}{PMF}&
\multirow{2}{*}{F1} &
Orig. &13.72&20.83&13.30&7.53&15.62&24.09&15.08&9.01&11.34&16.43&11.02&5.41\\
&&Fair &\textbf{14.01}&15.80&\textbf{13.90}&\textbf{1.90}&\textbf{15.84}&18.01&\textbf{15.71}&\textbf{2.30}&\textbf{11.35}&12.68&\textbf{11.27}&\textbf{1.41}\\
 \cline{2-15}
&\multirow{2}{*}{NDCG}&
Orig. &41.06&50.35&40.52&9.83&44.85&58.39&43.99&14.40&36.23&46.37&35.60&10.77\\
&&Fair &\textbf{41.28}&42.28&\textbf{41.22}&\textbf{1.06}&\textbf{44.98}&48.74&\textbf{44.74}&\textbf{4.00}&\textbf{36.27}&42.06&\textbf{35.91}&\textbf{6.15}\\ 
\hline
\multirow{4}{*}{NeuMF}&
\multirow{2}{*}{F1} &
Orig. &12.44&20.36&11.97&8.39&14.51&22.84&13.98&8.86&12.20&21.12&11.65&9.47\\
&&Fair &\textbf{12.81}&14.70&\textbf{12.70}&\textbf{2.00}&\textbf{14.82}&15.76&\textbf{14.76}&\textbf{1.00}&\textbf{12.36}&16.13&\textbf{12.13}&\textbf{4.00}\\
 \cline{2-15}
&\multirow{2}{*}{NDCG}&
Orig. &35.13&46.92&34.43&12.49&38.86&51.48&38.07&13.41&33.55&46.51&32.74&13.77\\
&&Fair &\textbf{35.31}&34.87&\textbf{35.34}&\textbf{0.47}&\textbf{38.98}&38.79&\textbf{39.00}&\textbf{0.21}&\textbf{33.66}&39.17&\textbf{33.32}&\textbf{5.85}\\ 
\hline
\multirow{4}{*}{STAMP}&
\multirow{2}{*}{F1} &
Orig. &12.75&19.04&12.38&6.66&14.41&21.75&13.96&7.79&12.86&20.02&12.42&7.60\\
&&Fair &\textbf{12.84}&13.98&\textbf{12.77}&\textbf{1.21}&\textbf{14.53}&16.31&\textbf{14.41}&\textbf{1.90}&\textbf{12.88}&16.46&\textbf{12.66}&\textbf{3.80}\\
 \cline{2-15}
&\multirow{2}{*}{NDCG}&
Orig. &35.54&44.27&35.02&9.25&38.16&48.39&37.51&10.88&35.83&46.57&35.17&11.40\\
&&Fair &\textbf{35.60}&37.31&\textbf{35.50}&\textbf{1.81}&\textbf{38.22}&41.76&\textbf{38.00}&\textbf{3.76}&\textbf{35.84}&42.27&\textbf{35.44}&\textbf{6.83}\\ 
 \bottomrule
\end{tabular}
\vspace{-10pt}
\end{table*}
%%%%%%%%%%%%%%%

\subsection{Main Results}
In this section, we show the performance of our re-ranking method on both of the recommendation quality and fairness effectiveness compared with traditional fairness-unaware recommendation algorithms.

Table~\ref{record}, Table~\ref{sum} and Table~\ref{max} show the main results on the three Amazon datasets about dividing user groups based on their number of interactions, total consumption and maximum price respectively. The overall performance, advantaged performance and disadvantaged performance are calculated on the whole test set, the group of advantaged users in the test set, and the group of disadvantaged users in the test set respectively. UGF is computed as Equation~\ref{fair} to evaluate the difference of recommendation quality ($\text{NDCG} @ K$ or $\mathrm{F}_{1}$) between the advantaged and disadvantaged groups. We set the upper bound of the constraint $\epsilon$ to the half of the metric differences between two groups of the fairness-unaware baseline. The original results in the table are from baselines, and the fair results in the table are from our model.

Comparing advantaged and disadvantaged groups under four baselines, we can find that there is a big difference in recommendation performance between the two groups. Take the results of NeuMF on Grocery as an example, in Table~\ref{record}, the difference of $\text{NDCG} @ 10$ between two groups is 32.21\%, and the difference of $\mathrm{F}_{1}@ 10$ between two groups is 23.51\%; in Table~\ref{sum}, the difference of $\text{NDCG} @ 10$ is 27.43\%, and the difference of $\mathrm{F}_{1}@ 10$ is 20.34\%; in Table~\ref{max}, the difference of $\text{NDCG} @ 10$ is 13.41\%, and the difference of $\mathrm{F}_{1}@ 10$ is 8.86\%.
Such disparity could be caused by the nature of collaborative filtering. In other words, the advantaged users may dominate the learning algorithm, and thus the disadvantaged users are more likely to receive biased recommendations due to their insufficient training data, which results in extremely unfair treatments by the system.

We can see from the three tables that our re-ranking method has the ability to significantly reduce the fairness disparity as well as improve the overall recommendation performance of all baselines. For example, also from results of NeuMF on Grocery in Table.\ref{record}, fair-NeuMF improves the overall $\text{NDCG} @ 10$ from 38.86\% to 40.09\%, and improves the overall $\mathrm{F}_{1}@ 10$ from 14.51\% to 15.48\%, as well as reduces the difference between $\text{NDCG} @ 10$ from 32.21\% to 10.85\%, and  reduces the difference between $\mathrm{F}_{1}@ 10$ from 23.51\% to 5.82\%. What's more,  the performance of disadvantaged groups has also been improved due to the fairness constraint. For example, $\text{NDCG} @ 10$ is improved from 36.13\% to 39.17\%, and $\mathrm{F}_{1}@ 10$ from 12.51\% to 14.99\%. However, the performance of advantaged users is reduced to satisfy our fairness constraint. Although we sacrifice some of the average recommendation performance of the advantaged users, the constraint that decreases the disparity between two groups substantially improves the performance of the disadvantaged group, which accounts for much more users than the advantaged users. The total performance compromise of the advantaged users is much smaller than the total improvement of the disadvantaged users, which is the reason why the overall performance gets boosted.

Among the three grouping methods, we see that the fairness disparity of dividing users based on their maximum purchase price is not as significant as dividing groups according to their number of records or total consumption. This could be a possible reason that limits the ability of performance improvement of our re-ranking method. The less significant fairness disparity of the max price grouping method may lie in the following two aspects. On the one hand, although users who interact more actively with platforms tend to have a greater consumption range, using the maximum price to capture users' activity may be less informative than the other two methods, since it cannot properly capture the accumulative influence in the long term. On the other hand, Figure \ref{distribution} in Section \ref{sec:motivation} shows the user distribution under three grouping methods. We can see that limited by the maximum price of items in each dataset, the difference of user distributions between advantaged and disadvantaged groups under the maximum price method is not as obvious as the other two methods, thus resulting in less significant unfair treatment by the systems under this grouping method. However, the unfair treatment of recommendation systems is still obvious under max price grouping method, which reflects its capability to capture user activity level. 

Overall, the experiments show that our re-ranking algorithm can not only shrink the fairness disparity between the two groups of users, but also provide better overall recommendation results than the baseline methods. With these more fairer recommendation lists, those disadvantaged users who account for a large proportion of the user community can get more benefits.

\vspace{-2pt}
\subsection{Ablation Study}
In different scenarios, the definition of fairness and the strictness of fairness requirement may be different. From Definition \ref{fair}, we know that the smaller $\epsilon$ is, the fairer our model will be. However, the excessive pursuit of fairness sometimes is unnecessary and will result in a great impact on the recommendation performance. As shown in the main results we presented above, to achieve fairer performance, the average recommendation quality of the advantaged users could be sacrificed, since their original scores were so high that the scores of disadvantaged users could not approach them through directly re-ranking the recommendation lists based on their own preference scores. Therefore, we are interested in how the value of $\epsilon$, which can evaluate fairness between two groups, will affect the recommendation quality of different groups.

\begin{figure}[t]
        \centering
        \begin{subfigure}[b]{0.235\textwidth}
            \centering
            \resizebox{0.9\textwidth}{!}{
                \begin{tikzpicture}
                    \pgfplotsset{
                        scale only axis,
                         label style={font=\Huge},
                         tick label style={font=\huge},
                        xmin=0, xmax=0.8
                    }
                    \begin{axis}[
                    %   axis y line*=left,
                      ymin=8, ymax=35,
                      xlabel={$1/\epsilon$},
                      ylabel={F1@10},
                      xticklabels={0,{1/17.6},{},{1/11.7},{},{1/5.8},{},{1/2.9},{},{$\epsilon=0$}},
                      legend pos=north east,
                      ymajorgrids=true,
                     legend style={font=\fontsize{10}{5}\selectfont}
                    ]
                    \addplot[mark=square,blue, line width=1.5pt]
                     coordinates {
                      (0,14.69)
                        (0.2,15.07)
                        (0.4,15.48)
                        (0.6,15.62)
                        (0.8, 15.70)
                    }; 
                    \addlegendentry{Overall}
                    \addplot[mark=square,green, line width=1.5pt]
                     coordinates {
                        (0,30.80)
                        (0.2,25.78)
                        (0.4,20.81)
                        (0.6,18.33)
                        (0.8,15.83)
                    };
                    \addlegendentry{Advantaged}
                    \addplot[mark=square,red, line width=1.5pt]
                     coordinates {
                        (0,13.20)
                        (0.2,14.07)
                        (0.4,14.99)
                        (0.6,15.37)
                        (0.8,15.69)
                    };
                    \label{plot_boolean_logic}
                    \addlegendentry{Disadvantaged}
                    \end{axis}
                \end{tikzpicture}
            }
            \caption[Network2]%
            {{\small interactions}}    
            \label{fig:mean and std of net14}
        \end{subfigure}
        \hfill
        \begin{subfigure}[b]{0.235\textwidth}  
            \centering 
            \resizebox{0.9\textwidth}{!}{
                \begin{tikzpicture}
                    \pgfplotsset{
                        scale only axis,
                         label style={font=\Huge},
                         tick label style={font=\huge},
                        xmin=0, xmax=0.8
                    }
                    \begin{axis}[
                    %   axis y line*=left,
                      ymin=31, ymax=68,
                      xlabel={$1/\epsilon$},
                      ylabel={NDCG@10},
                      xticklabels={0,{1/17.6},{},{1/11.7},{},{1/5.8},{},{1/2.9},{},{$\epsilon=0$}},
                      legend pos=north east,
                      ymajorgrids=true,
                      legend style={font=\fontsize{10}{5}\selectfont}
                    ]
                   \addplot[mark=square,blue, line width=1.5pt]
                     coordinates {
                      (0,39.29)
                        (0.2,39.68)
                        (0.4,40.09)
                        (0.6,40.25)
                        (0.8,40.22)
                    
                    }; 
                    \addlegendentry{Overall}
                    \addplot[mark=square,green, line width=1.5pt]
                     coordinates {
                        (0,64.36)
                        (0.2,57.24)
                        (0.4,50.02)
                        (0.6,46.77)
                        (0.8,42.56)
                    };
                     \addlegendentry{Advantaged}
                    \addplot[mark=square,red, line width=1.5pt]
                     coordinates {
                        (0,36.96)
                        (0.2,38.06)
                        (0.4,39.17)
                        (0.6,39.65)
                        (0.8,40.00)
                    };
                    \addlegendentry{Disadvantaged}
                    \label{plot_boolean_logic}
                    \end{axis}
                \end{tikzpicture}
            }
            \caption[]%
            {{\small interactions}}    
            \label{fig:mean and std of net24}
        \end{subfigure}
        \vskip\baselineskip
        \begin{subfigure}[b]{0.235\textwidth}
            \centering
            \resizebox{0.9\textwidth}{!}{
                \begin{tikzpicture}
                    \pgfplotsset{
                        scale only axis,
                         label style={font=\Huge},
                         tick label style={font=\huge},
                        xmin=0, xmax=0.8
                    }
                    \begin{axis}[
                    %   axis y line*=left,
                      ymin=8, ymax=35,
                      xlabel={$1/\epsilon$},
                      ylabel={F1@10},
                      xticklabels={0,{1/15.3},{},{1/10},{},{1/5},{},{1/2},{},{$\epsilon=0$}},
                      legend pos=north east,
                      ymajorgrids=true,
                      legend style={font=\fontsize{10}{5}\selectfont}
                    ]
                    \addplot[mark=square,blue, line width=1.5pt]
                     coordinates {
                      (0,14.63)
                        (0.2,14.85)
                        (0.4,15.24)
                        (0.6,15.49)
                        (0.8,15.62)
                    }; 
                     \addlegendentry{Overall}
                    \addplot[mark=square,green, line width=1.5pt]
                     coordinates {
                        (0,28.66)
                        (0.2,24.02)
                        (0.4,19.84)
                        (0.6,17.34)
                        (0.8,15.68)
                    };
                     \addlegendentry{Advantaged}
                    \addplot[mark=square,red, line width=1.5pt]
                     coordinates {
                        (0,13.36)
                        (0.2,14.02)
                        (0.4,14.83)
                        (0.6,15.32)
                        (0.8,15.62)
                    };
                     \addlegendentry{Disadvantaged}
                    \label{plot_boolean_logic}
                    % \addlegendentry{plot 1}
                    \end{axis}
                \end{tikzpicture}
            }
            \caption[Network2]%
            {{\small total consumption}}    
            \label{fig:mean and std of net14}
        \end{subfigure}
        \hfill
        \begin{subfigure}[b]{0.235\textwidth}  
            \centering 
            \resizebox{0.9\textwidth}{!}{
                \begin{tikzpicture}
                    \pgfplotsset{
                        scale only axis,
                         label style={font=\Huge},
                         tick label style={font=\huge},
                        xmin=0, xmax=0.8
                    }
                    \begin{axis}[
                    %   axis y line*=left,
                      ymin=31, ymax=68,
                      xlabel={$1/\epsilon$},
                      ylabel={NDCG@10},
                      xticklabels={0,{1/15.3},{},{1/10},{},{1/5},{},{1/2},{},{$\epsilon=0$}},
                      legend pos=north east,
                      ymajorgrids=true,
                      legend style={font=\fontsize{10}{5}\selectfont}
                    ]
                   \addplot[mark=square,blue, line width=1.5pt]
                     coordinates {
                      (0,39.13)
                        (0.2,39.24)
                        (0.4,39.60)
                        (0.6,39.84)
                        (0.8,40.03)
                    }; 
                     \addlegendentry{Overall}
                    \addplot[mark=square,green, line width=1.5pt]
                     coordinates {
                        (0,60.01)
                        (0.2,52.42)
                        (0.4,45.31)
                        (0.6,41.55)
                        (0.8,39.79)
                    };
                     \addlegendentry{Advantaged}
                    \addplot[mark=square,red, line width=1.5pt]
                     coordinates {
                        (0,37.23)
                        (0.2,38.05)
                        (0.4,39.08)
                        (0.6,39.69)
                        (0.8,40.05)
                    };
                     \addlegendentry{Disadvantaged}
                    \label{plot_boolean_logic}
                    \end{axis}
                \end{tikzpicture}
            }
            \caption[]%
            {{\small total consumption}}    
            \label{fig:mean and std of net24}
        \end{subfigure}
         \vskip\baselineskip
        \begin{subfigure}[b]{0.235\textwidth}
            \centering
            \resizebox{0.9\textwidth}{!}{
                \begin{tikzpicture}
                    \pgfplotsset{
                        scale only axis,
                         label style={font=\Huge},
                         tick label style={font=\huge},
                        xmin=0, xmax=0.8
                    }
                    \begin{axis}[
                    %   axis y line*=left,
                      ymin=13, ymax=21.5,
                      xlabel={$1/\epsilon$},
                      ylabel={F1@10},
                      xticklabels={0,{1/6.6},{},{1/4},{},{1/2},{},{1},{},{$\epsilon=0$}},
                      legend pos=north east,
                      ymajorgrids=true,
                      legend style={font=\fontsize{10}{5}\selectfont}
                    ]
                    \addplot[mark=square,blue, line width=1.5pt]
                     coordinates {
                        (0,14.53)
                        (0.2,14.67)
                        (0.4,14.79)
                        (0.6,14.82)
                        (0.8,14.93)
                    }; 
                     \addlegendentry{Overall}
                    \addplot[mark=square,green, line width=1.5pt]
                     coordinates {
                        (0,20.73)
                        (0.2,18.43)
                        (0.4,16.67)
                        (0.6,15.76)
                        (0.8,14.93)
                    };
                    \addlegendentry{Advantaged}
                    \addplot[mark=square,red, line width=1.5pt]
                     coordinates {
                        (0,14.13)
                        (0.2,14.43)
                        (0.4,14.67)
                        (0.6,14.76)
                        (0.8,14.93)
                    };
                    \addlegendentry{Disadvantaged}
                    \label{plot_boolean_logic}
                    % \addlegendentry{plot 1}
                    \end{axis}
                \end{tikzpicture}
            }
            \caption[Network2]%
            {{\small max price}}    
            \label{fig:mean and std of net14}
        \end{subfigure}
        \hfill
        \begin{subfigure}[b]{0.235\textwidth}  
            \centering 
            \resizebox{0.9\textwidth}{!}{
                \begin{tikzpicture}
                    \pgfplotsset{
                        scale only axis,
                         label style={font=\Huge},
                         tick label style={font=\huge},
                        xmin=0, xmax=0.8
                    }
                    \begin{axis}[
                    %   axis y line*=left,
                      ymin=35.5, ymax=50,
                      xlabel={$1/\epsilon$},
                      ylabel={NDCG@10},
                      xticklabels={0,{1/6.6},{},{1/4},{},{1/2},{},{1/1},{},{$\epsilon=0$}},
                      legend pos=north east,
                      ymajorgrids=true,
                      legend style={font=\fontsize{10}{5}\selectfont}
                    ]
                   \addplot[mark=square,blue, line width=1.5pt]
                     coordinates {
                      (0,38.86)
                        (0.2,38.97)
                        (0.4,38.99)
                        (0.6,38.98)
                        (0.8,39.07)
                    }; 
                    \addlegendentry{Overall}
                    \addplot[mark=square,green, line width=1.5pt]
                     coordinates {
                        (0,49.00)
                        (0.2,44.96)
                        (0.4,40.49)
                        (0.6,38.79)
                        (0.8,36.94)
                    };
                    \addlegendentry{Advantaged}
                    \addplot[mark=square,red, line width=1.5pt]
                     coordinates {
                        (0,38.22)
                        (0.2,38.59)
                        (0.4,38.89)
                        (0.6,39.00)
                        (0.8,39.21)
                    };
                    \addlegendentry{Disadvantaged}
                    \label{plot_boolean_logic}
                    \end{axis}
                \end{tikzpicture}
            }
            \caption[]%
            {{\small max price}}
            \label{fig:mean and std of net24}
        \end{subfigure}
        \caption[ The average and standard deviation of critical parameters ]
        {\small The metrics F$_1$@10 and NDCG@10 change with respect to the $\epsilon$ on Overall, Advantaged and Disadvantaged groups.} 
        \label{ablation}
        \vspace{-30pt}
\end{figure}
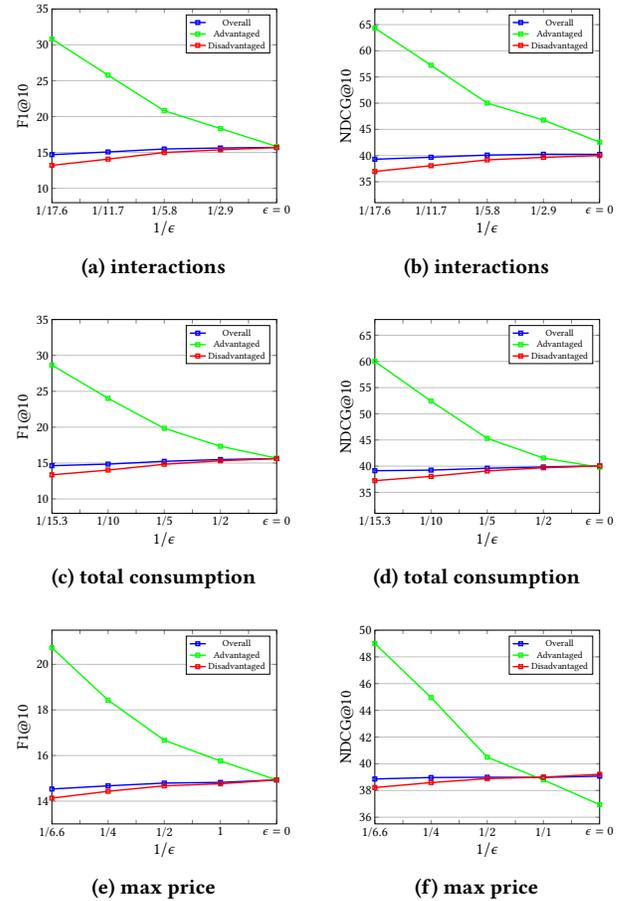

In this section, we study how the value of $\epsilon$ in Equation \ref{fair} can influence the performance of the overall, advantaged group and disadvantaged group. Take the performance of fair-NeuMF on the Grocery dataset as an example, Figure.\ref{ablation} shows how recommendation quality changes with the degree of relaxation of fairness requirements.

All results are under the fairness constraint of $\mathrm{F}_{1}@ 10$. From the figures, we can see that the more stringent the requirement of fairness is, the more performance reduction of the advantaged group, and the more performance improvement of overall and disadvantaged group. And the $\mathrm{F}_{1} @ 10$ of these three groups will be almost the same when we set $\epsilon = 0$. Although the results are generated under the constraint of $\mathrm{F}_{1} @ 10$, we can see that the recommendation quality on $\text{NDCG} @ 10$ also shows a similar trend in performance change. In Figure.\ref{ablation}(f), the performance of the advantaged group is even lower than the disadvantaged group when we set $\epsilon = 0$. Therefore, there may be a trade-off between pursuing fairness and reducing the sacrifice of the advantaged group under some scenarios, although we can still get improvements on the overall performance.

\section{Conclusions}
\label{sec:conclusions}
In this work, we study the problem of fairness in recommendation algorithms from the user perspective. We show that current recommendation algorithms will capture the data imbalance that lies on the user side, thus produces unfair treatment between different user groups.
% Specifically, we explore three methods to group users into advantaged and disadvantaged groups in commercial recommendation systems according to their activity level. 
We first conduct a data-driven observation analysis on three Amazon datasets with several shallow or deep recommendation algorithms, to show that users who interact more actively with platforms only account for a small proportion of users in data. However, the recommendation quality for these advantaged users is significantly higher than those disadvantaged users, which gives rise to unfair issues in recommender systems. The unfair treatment between different groups of users can also reduce the overall performance since the less active users are in the majority. We then quantify unfairness at the group level and provide a fairness constrained re-ranking method to mitigate the unfairness between advantaged and disadvantaged groups while maintaining the recommendation quality. Our extensive experiments show that our method can reduce the unfairness between advantaged and disadvantaged groups significantly, and also improve the overall recommendation quality through providing more satisfying recommendations to the majority of disadvantaged users. 

\section*{Acknowledgement}
We appreciate the reviews and suggestions of the reviewers. This work was supported in part by NSF IIS-1910154 and IIS-2007907. Any opinions, findings, conclusions or recommendations expressed in this material are those of the authors and do not necessarily reflect those of the sponsors.

\bibliographystyle{ACM-Reference-Format}
\balance
\bibliography{sample-base}

%%
%% If your work has an appendix, this is the place to put it.
\appendix

\end{document}